\documentclass[cits]{PoS}

\newcommand{\nn}{\nonumber}

\title{Estimate of the systematic error in tau decay due to duality violations}

\ShortTitle{Duality Violations}

\author{Oscar Cat\`a\\
       INFN, Laboratori Nazionali di Frascati, Via E. Fermi 40, I-00044 Frascati, Italy\\
       E-mail: \email{ocata@ifae.es}}

       \author{Maarten Golterman\\
     Department of Physics and Astronomy, San Francisco State
University, 1600 Holloway Ave, San Francisco, CA 94132, USA\\
       E-mail: \email{maarten@stars.sfsu.edu}}

\author{\speaker{Santiago Peris}\thanks{This work was supported in part by
CICYT-FEDER-FPA2005-02211, SGR2005-00916, the Spanish Consolider-Ingenio 2010 Program CPAN (CSD2007-00042), by the EU Contract No. MRTN-CT-2006-035482, ``FLAVIAnet,'' and by the US Department of Energy.
}\\
        Grup de F{\'\i}sica Te{\`o}rica and IFAE, UAB, E-08193 Bellaterra, Barcelona, Spain\\
        E-mail: \email{peris@ifae.es}}

\abstract{With the help of a physically motivated ansatz, which is fitted to the data, we estimate the size of possible duality violations in hadronic tau decay. The result is that the uncertainty associated with these violations could impact the $\alpha_s$ determination from the total decay width at a level which we estimate to be $\delta\alpha_s(m_{\tau})\sim 0.003-0.010$. Consequently, it cannot be neglected.}

\FullConference{8th Conference Quark Confinement and the Hadron Spectrum \\
         September 1-6 2008\\
         Mainz, Germany}

\begin{document}

\section{Introduction}

\begin{figure}
\centering
\includegraphics[width=0.35\textwidth]{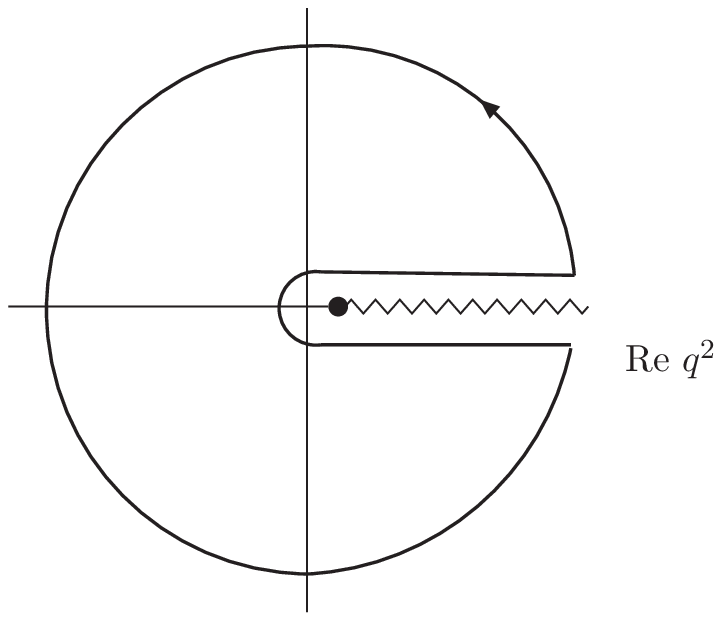}
\hspace{1cm}
\includegraphics[width=0.35\textwidth]{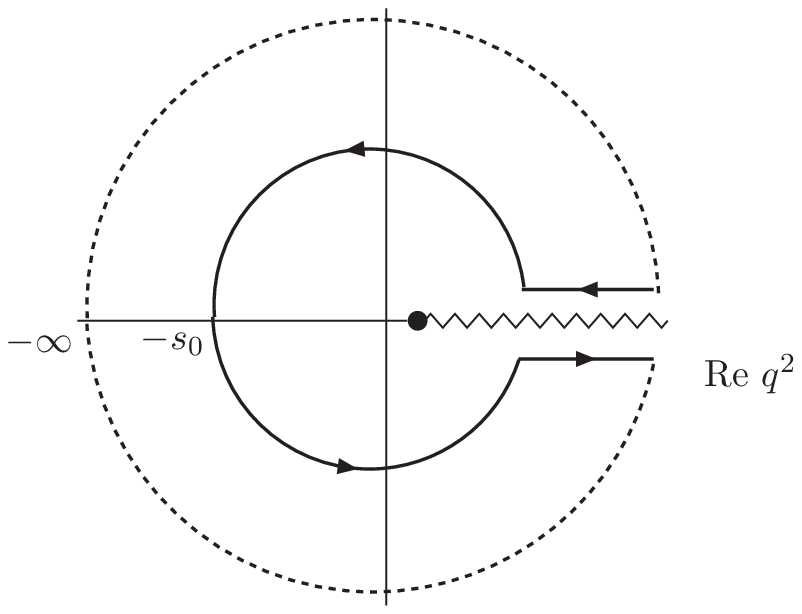}
\caption{ Left panel: Analytic structure of $\Pi_{V,A}(q^2)$ in the complex $q^2$ plane. The solid curve shows the contour used in Eq.~(1.3). Right panel: Contour used in the derivation of Eq.~(2.3). With the assumed exponential decay, $\Delta_{V,A}(q^2)$ vanish on the dashed circle with infinite radius.}\label{cauchyfig}
\end{figure}

After the seminal work of Ref. \cite{BNP}, hadronic tau decay has emerged as a very interesting physical process for studying QCD. Recent analyses  find for the coupling constant $\alpha_s$ the following values:
\begin{eqnarray}\label{alphas}
\alpha_s(m_{\tau}^2)&=0.344\pm 0.005_{\mathrm{exp}}\pm 0.007_{\mathrm{th}}\ [2] \ , \quad \alpha_s(m_{\tau}^2)&=0.332\pm 0.005_{\mathrm{exp}}\pm 0.015_{\mathrm{th}}\ [3] \ , \nonumber \\
\alpha_s(m_{\tau}^2)&=0.321\pm 0.005_{\mathrm{exp}}\pm 0.012_{\mathrm{th}}\ [4]\ , \quad
\alpha_s(m_{\tau}^2)&=0.316\pm 0.003_{\mathrm{exp}}\pm 0.005_{\mathrm{th}}\ [5]\  .
\end{eqnarray}
 These results show an unprecedented level of precision.  It is, therefore, very important to be able to assess all the  uncertainties involved in these analyses.

Furthermore, Ref. \cite{Davier}, after performing a very complete study of the vector and axial spectral functions extracted from tau decay, obtains for the gluon condensate a \emph{different} number  depending on which channel is used:
\begin{equation} \label{gluonVA}
\frac{\alpha_s}{\pi}\langle GG\rangle\Big|_{\mathrm{Vector}}\!\!\!\!\!\!\!\!\!\!\!\!=(-0.8\pm0.4)\times
10^{-2}\ {\mathrm{GeV}}^4\ , \quad
\frac{\alpha_s}{\pi}\langle
GG\rangle\Big|_{\mathrm{Axial}}\!\!\!\!\!\!\!\!\!\!\!=(-2.2\pm0.4)\times 10^{-2}\ {\mathrm{GeV}}^4\ ,\nn
\end{equation}
which, if taken at face value, would reflect an inconsistency in the Operator Product Expansion (OPE), as the gluon condensate should be the same in both channels \cite{SVZ}. Further determinations of the OPE condensates may be found in Refs. \cite{condensates:theworks}.

In the present situation we think it is time to reassess the level of precision one may expect from the OPE in tau decay \cite{Thebomb}. In fact, all the analyses involve an integration of the OPE over the physical cut where, in fact, this expansion is  expected \emph{not} to converge. The central equation for this discussion is \cite{BNP, Shankar}:
\begin{equation}\label{cauchy1}
\int_0^{s_0}\,ds\,\, P(s) \,\frac{1}{\pi}{\mathrm{Im}}\,\Pi_{V,A}(s)=-\frac{1}{2\pi i}\oint_{|q^2|=s_0}\,dq^2
\,P(q^2)\, \Pi^{\mathrm{OPE}}_{V,A} (q^2)\ ,
\end{equation}
where  $P(s)$ is any polynomial which may be chosen at one's convenience \cite{Pich}, and the contour of integration is shown on the left panel of Fig. \ref{cauchyfig} (note how the contour $|q^2|=s_0$ touches the physical cut).

Eq. (\ref{cauchy1}) allows one to compare the OPE (on the right hand side) with the experimental data (on the left hand side). From this comparison,  the extraction of $\alpha_s$ and the condensates follow. However, if the  OPE is not a good approximation on the physical cut, this extraction will have an error one ought to be able to estimate. Even though the polynomial $P(s)$ may be chosen to have a zero on this cut, this does not guarantee that this error will vanish. This type of error is commonly referred to as a Duality Violation (DV).

\begin{figure}
\centering
\includegraphics[width=2.5in]{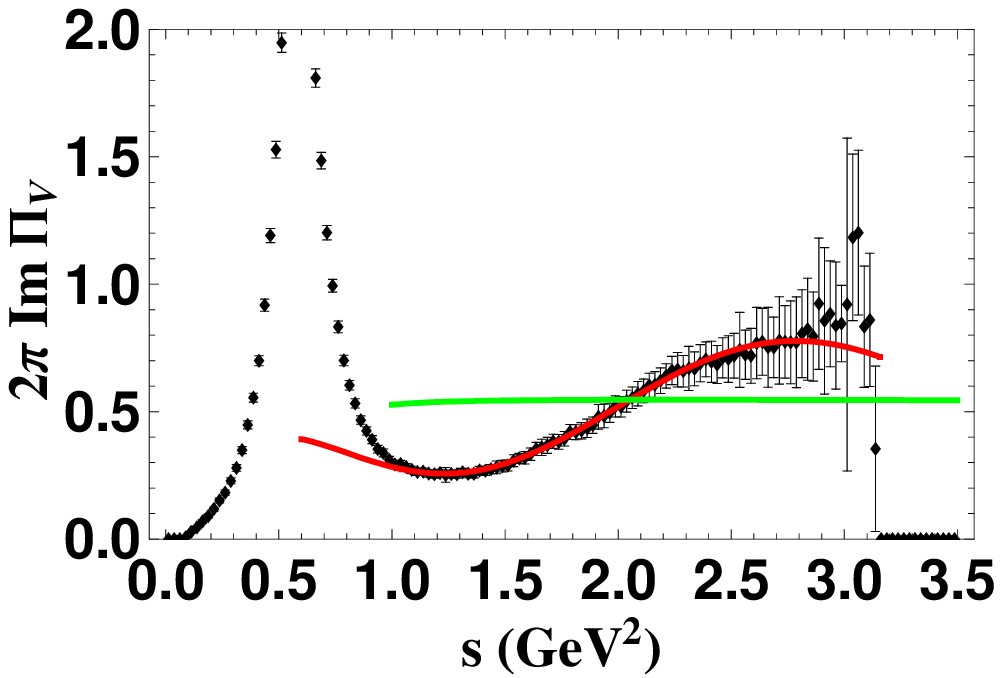}
\hspace{.2cm}
\includegraphics[width=2.5in]{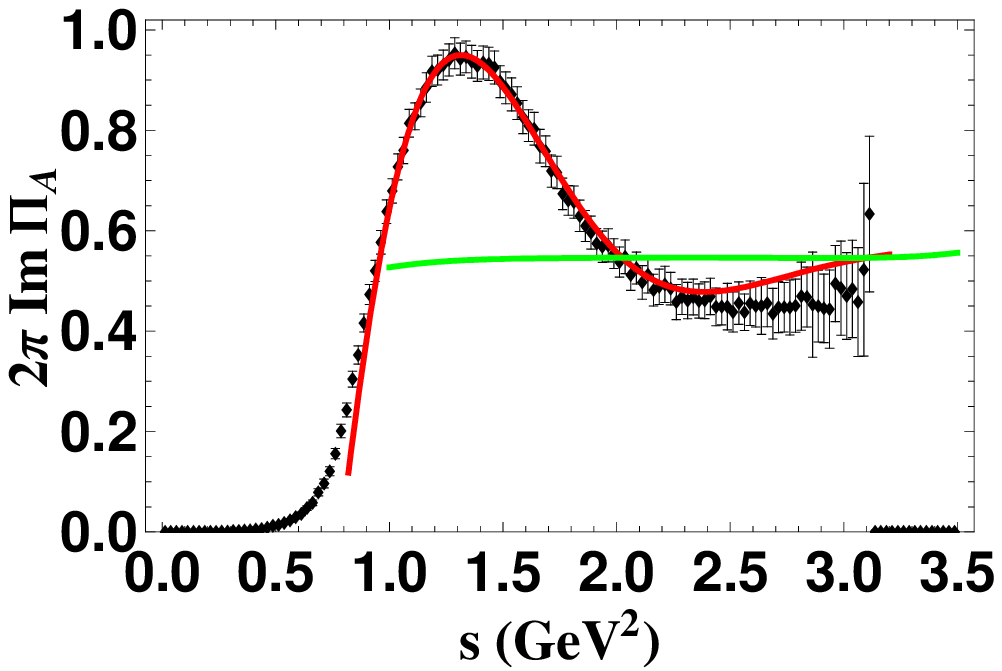}
\includegraphics[width=2.5in]{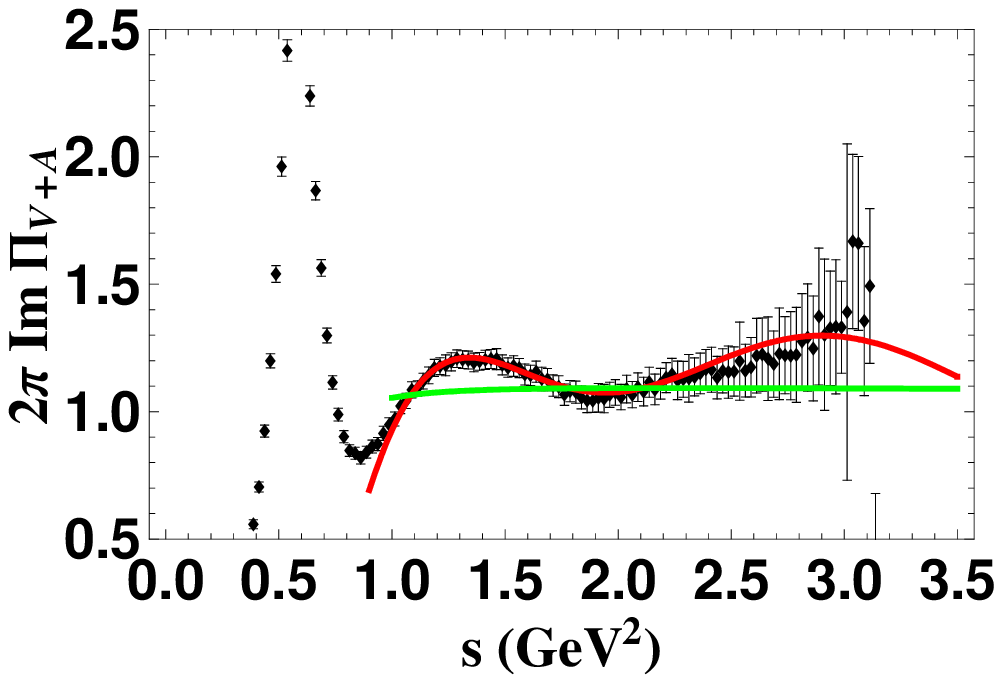}
\hspace{.2cm}
\includegraphics[width=2.5in]{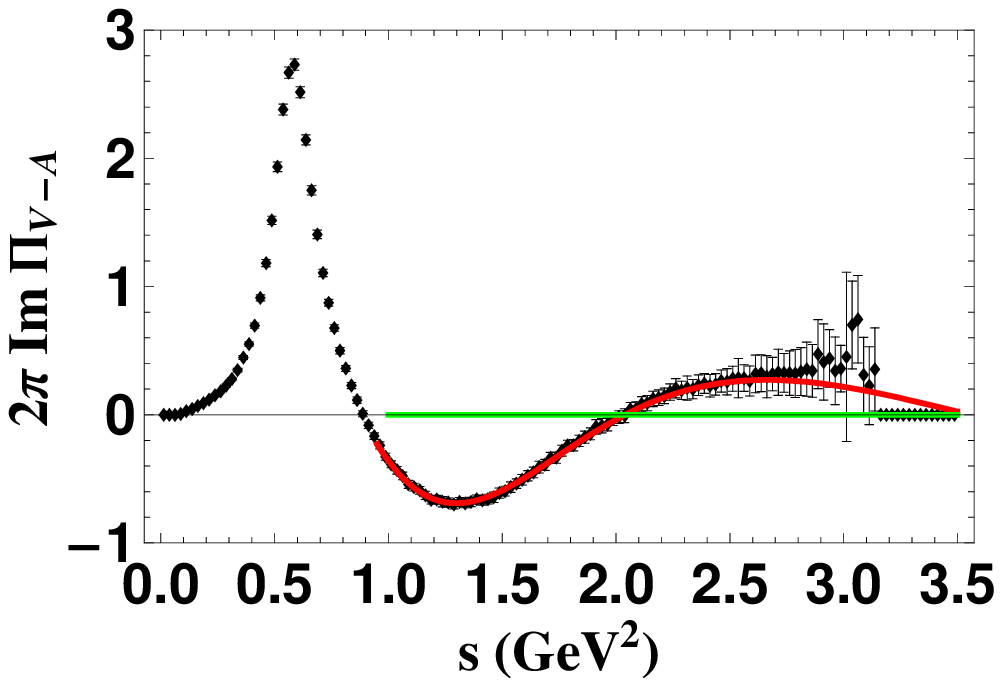}
\caption{Different combinations of spectral functions ($V$, $A$, and $V\pm A$)
(data points in black) compared with Fixed-Order perturbation theory (green flat line) and the result of our fit to Eq.~(2.4) (red curve).}\label{aleph-fig}
\end{figure}

\section{An ansatz for Duality Violations}

 Assuming that the OPE is an asymptotic expansion, one may expect that the difference
\begin{equation}\label{dv1}
    \Delta_{V,A}(q^2)=\Pi_{V,A} (q^2)-\Pi^{ \mathrm{OPE}}_{V,A} (q^2)  \ .
\end{equation}
will be exponentially suppressed at $|q^2|=\infty$. In this case,  use of Cauchy's theorem on the contour depicted on the right  panel in Fig. \ref{cauchyfig} allows one to obtain the correction to Eq. (\ref{cauchy1}) as
\begin{equation}\label{cauchy3}
\int_0^{s_0}\,ds\,\, P(s) \,\frac{1}{\pi}{\mathrm{Im}}\,\Pi_{V,A}(s)=-\frac{1}{2\pi i}\oint_{|q^2|=s_0}\,dq^2
\,P(q^2)\, \Pi^{OPE}_{V,A} (q^2)+ {\cal{D}}_{V,A}^{[P]}(s_0) \  .
\end{equation}
where \cite{CGP1,CGP2}
\begin{equation}\label{ourbaby}
    {\cal{D}}_{V,A}^{[P]}(s_0) =- \int_{s_0}^{\infty} ds\ P(s)\  \frac{1}{\pi} \mathrm{Im}\Delta_{V,A}(s)\ .
\end{equation}
encapsulates all the DVs. The function $\mathrm{Im}\Delta_{V,A}(s)$ in the previous equation requires knowledge of the spectrum up to infinite energy and is, in principle, unknown.

However, if indeed the OPE is an asymptotic expansion at $|q^2|=\infty$ ,  $\mathrm{Im}\Delta_{V,A}(s)$ in Eq. (\ref{ourbaby}) will exhibit some type of exponential falloff in $s$, at large values of the momentum. If, furthermore, we also assume that the spectrum follows some type of periodicity, such as the one found in the daughter trajectories in Regge theory,  it becomes natural to combine an oscillatory behavior with the  exponential falloff and assume the simple form
\begin{equation}\label{ansatz}
   \frac{1}{\pi} \mathrm{Im}\Delta_{V,A}(s)= \kappa_{V,A} \ e^{-\gamma_{V,A} s}\ \sin\left(\alpha_{V,A} +\beta_{V,A} s \right)\ ,
\end{equation}
as an ansatz to fit to the experimental data. We note that a model exists \cite{Shifman, Bigi, CGP1, CGP2} that realizes all these expected features of the OPE, and for which Eq. (\ref{ansatz}) gives the behavior at large values of $s$.

\section{Fitting the ansatz to tau data}

Armed with the ansatz (\ref{ansatz}) we have fitted the spectral functions in tau decay in the window $1.1$ GeV$^2 \leq s \leq m_{\tau}^2$. This window has been chosen because going below the lower limit in this window increases the $\chi^2/dof$ very sharply, while increasing the lower limit only increases the error of the fit (as the data are less precise), even up to a point where the fit ceases to be meaningful. The results of the fits we obtained are:
\begin{eqnarray}\label{fitV}
   \kappa_V = 0.018\pm 0.004 \qquad \qquad &,& \quad  \kappa_A = \ 0.20\pm 0.06 \quad \quad \ , \nn \\
  \gamma_V = 0.15\pm 0.15  \quad  \mathrm{GeV}^{-2}\quad &,& \quad   \gamma_A = \  1.7\pm 0.2  \quad  \mathrm{GeV}^{-2}\ , \nn\\
   \alpha_V = 2.2\pm 0.3 \qquad \qquad \qquad &,& \quad    \alpha_A = -0.4\pm 0.1 \quad \quad , \nn \\
 \beta_V = 2.0\pm 0.1  \quad \mathrm{GeV}^{-2}\qquad &,& \quad  \beta_A = -3.0\pm 0.1 \quad\   \mathrm{GeV}^{-2}\ , \nn \\
   \frac{\chi^2}{dof} = \frac{10}{79}\simeq 0.13\quad \qquad \qquad &,& \quad  \frac{\chi^2}{dof} = \frac{17}{78}\simeq 0.22\quad .
\end{eqnarray}
The contribution from the OPE to the spectral functions is negligible \cite{Thebomb}.
As one can see in Fig. \ref{aleph-fig}, and from the values of the $\chi^2$'s, the fits are very good. We remark that this is also true for the fit to the $V+A$ spectral function, which shows that there is no reason why this combination of spectral functions should be protected from DVs, unlike what is sometimes assumed in the literature. In fact, because the exponent $\gamma_A$ is much larger than $\gamma_V$, DVs in the $A$ channel are much suppressed compared to the $V$ channel, so that $V+A$ has the same DVs as $V$ alone.

Since, within the ansatz (\ref{ansatz}), DVs are sizeable only in the $V$ channel, it makes sense to ask whether $e^+e^-$ data modify the above conclusions. Although answering this requires new assumptions, the net result is that DVs tend to be decreased somewhat, but not to a level at which they become negligible. See the discussion in Ref. \cite{Thebomb}.

\section{Impact of Duality Violations on $\alpha_s$}

Choosing the right polynomial $P(s)$ in Eq. (\ref{cauchy3}) one may obtain the tau decay width on the left hand side of this equation as a function of the OPE parameters and the contribution from DVs (\ref{cauchy3},\ref{ourbaby}) in terms of the fit parameters (\ref{fitV}). Using the total tau decay width, $R_{\tau}$ \cite{BNP}, plus the conservative estimate for the condensates made in \cite{Jamin}, this leads to a shift in $\alpha_s$ which is \cite{Thebomb}
\begin{equation}\label{shift}
  \delta \alpha_s(m_{\tau}) \sim  0.003-0.010 \ .
 \end{equation}
The spread of values in (\ref{shift}) reflects the error of the fits and is insensitive to  whether one uses the Contour-Improved or the Fixed-Order prescriptions for perturbation theory \cite{Jamin}. In Ref. \cite{Davier} an estimate of DVs was also made, with the conclusion that DVs were negligible. However, as far as we understand, no attempt at a detailed fit to spectral data was made there. When the fit is made, our conclusion is that the systematic error due to DVs is not negligible and may even get to be of the order of all the other systematic errors taken together. We refer to Ref.~\cite{Thebomb} for details and further discussions.

\vspace{.5cm}
We thank M. Jamin for his collaboration at different stages of this work, as well as for discussions. We also thank M. Davier, S. Descotes-Genon, A. H\"{o}cker, L. Lyons, B. Malaescu, K. Maltman, R. Miquel, L. Mir, A. Pich, E. de Rafael and G. Venanzoni for correspondence and/or discussions.

\end{document}